# Title: Droplet's Acoustics

Raphael Dahan, Leopoldo L. Martin, Tal Carmon

*Technion - Israel Institute of Technology, 32000 Haifa, Israel*

Abstract:

**Contrary to their capillary resonances (Rayleigh, 1879) and their optical resonances (Ashkin, 1977), droplets' acoustical resonances were rarely considered. Here we experimentally excite, for the first time, the acoustical resonances of a droplet that relies on sound instead of capillary waves. Droplet's vibrations at 37 MHz rates and quality factor of 80 are optically excited and interrogated at an optical threshold of 68 μW. Our vibrations span a spectral band that is 1000 times higher when compared with drop's previously-studied capillary vibration.**

Current excitation schemes in droplets could access their capillary shape oscillations[1] at kHz rates but not their acoustic ones. Here we use a MHz excitation rate, which provides droplet's acoustical modes where compression restores shape toward mechanical equilibrium. This is in contrast with capillary modes where surface tension pushes the drop back toward equilibrium. The calculated capillary and mechanical modes of a droplet are shown in Figure 1 where one can see the uniform pressure of the capillary mode, Side-by-side with the varying pressure of the acoustical mode. In this work we distinguish ourselves from previous studies in capillary resonance of drops (Figure **1a**) by going to their acoustic modes (Figure **1b**).

The analytical solution for the capillary eigenfrequency[1] of a spherical droplet in air is given by

$$(1)\ f = \sqrt{\frac{2\gamma}{\pi^2 \rho r^3}}$$

here $\rho$ is the liquid density, $\gamma$ is the surface tension at the air-liquid interface, and $r$ is the droplet radius. This is the natural frequency of the eigenfunction that is shown in Figure 1**a**.

Differently, the analytical solution for the mechanical eigenfrequency [2] of such a spherical droplet is given by

$$(2)\ f = 1.3\sqrt{\frac{B}{\pi^2 \rho r^2}}$$

Where $B$ is the bulk modulus of the medium, defined as its volume change per unit of pressure.

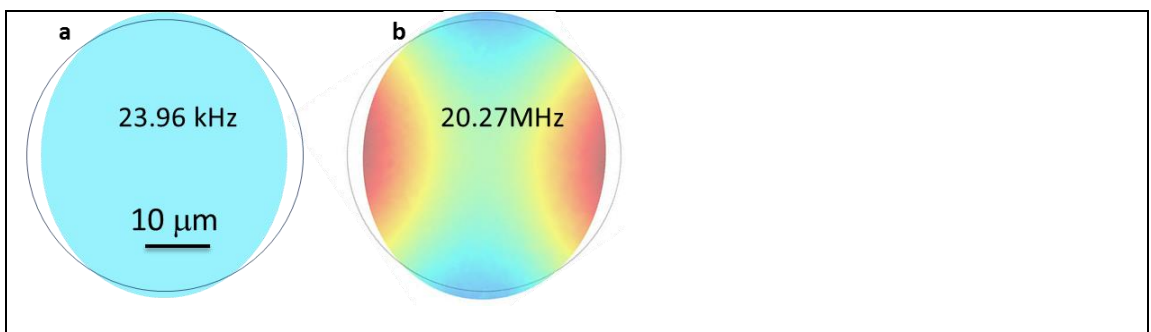

**Figure 1: Theory of droplet's modes**: comparison between a capillary mode **(a)** and its acoustical correspondent **(b)** of a liquid sphere in air. The outer circle represents the boundary of the sphere at equilibrium state and the color represents pressure. The liquid is Silicone oil as in the experiments here.

Interestingly, these two equations are similar except for changing B with $\gamma/r$.

For a typical liquid, this relation between the equations suggests that the mechanical modes of a droplet are a thousand times faster than their capillary ones. Accordingly, fast rates are needed in order to excite the acoustical modes of a droplet. Such fast rates can be achieved using the forces of light. In this regard, optomechanical [3-11] excitation schemes act quick [12], and can even excite vibrations at 10 GHz rates[13].A bridge between optomechanics and optofluidics[14-17] was recently established by demonstrating an optomechanically oscillating water-containing pipette [18,19]. Therefore, it is natural to ask if one can take a droplet and optically excite and interrogate its mechanical mode similarly to what was done with solid devices. When optically exciting the mechanical mode of a drop, one should first excite its optical mode. Optical modes of droplets were excited in levitating[20] and falling[21] droplets in experiments that mark the first days of micro cavity research. These early works were followed by excitation of optical resonances in droplets that were held by a stem [22] or laying on a hydrophobic surface[23]. As for the droplet stability, micro droplets possess a surprising stability. In essence, surface tension that holds 40 μm diameter droplets such as in our experiments is 6120 times stronger than gravity. This is calculated by the Eötvös number[24], a dimensionless number that represents the ratio between interfacial tension and gravity. This large Eötvös number that is typical to small-radii curvatures suggests that our droplets might withstand high accelerations of several Gs to enable liquid-walled micro optomechanics.

We start by fabricating a droplet resonator from Xiameter® silicone oil. This liquid was chosen since it combines high optical transmission together with a low evaporation rate. The droplet resonator is made by dipping a silica micro cylinder into the oil vessel. This cylinder is terminated with a ball shape on its end to eliminate droplet motion along the cylinder. When pulled out, the silica stem contains a micro droplet in its end as one can see in Figure 2.

Our experimental setup relies on a tunable laser diode, centered at 770 nm, that is evanescently coupled to the droplet's resonance via a tapered fiber [22,23,25]. Light is coupled out of the resonator through the other side of this taper into a fast photodetector in order to measure the optical transmission through the resonator.

In our first experiment we optically characterize our droplet resonators by measuring its quality factor. The droplet's optical quality factor is obtained using a ring down measurement [26] where the cavity is charged with light, followed by measuring the resonance decay time. As common in such measurements [26], while laser frequency and resonance frequency overlap, the cavity is charged with light. Right afterward, the resonance separates from the pump and the light that was stored in the cavity exponentially discharges. The discharging light interferes constructively and destructively with the drifting laser and hence the oscillation (Figure **2**). Ring down is preferred here over linewidth measurement, for measuring cavity Q, in order to avoid thermal broadening and narrowing effects[27]. We fit our experimentally measured transmission (Figure 2 red dots) to the rate equation model (blue line) for the optical resonator as appears in Gorodetsky et al [28]. Fitting the maxima and minima to an exponential decay rate we measured a lifetime of $69 \pm 5(ns)$ (inset of Figure **2**) corresponding to an optical quality factor of $Q = (8.5 \pm 0.5)10^7$ and a $(2.5 \pm 0.2)10^5$ finesse.

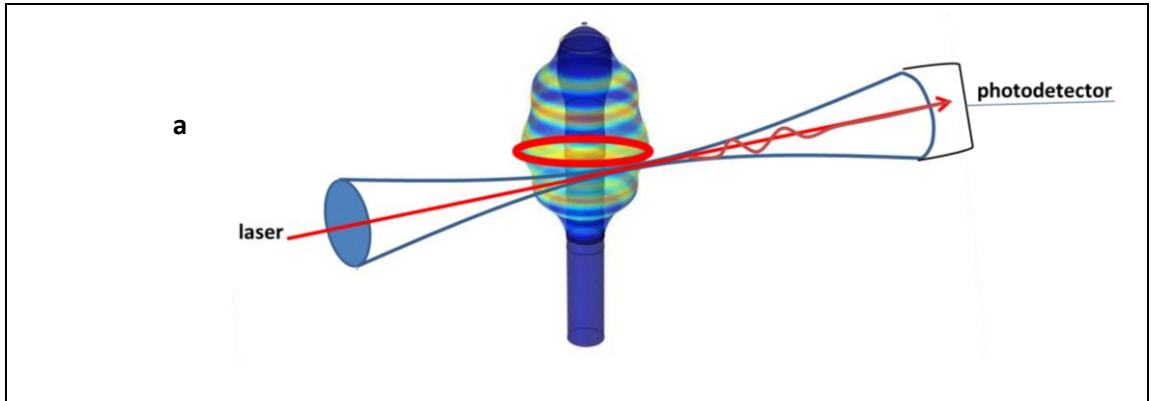

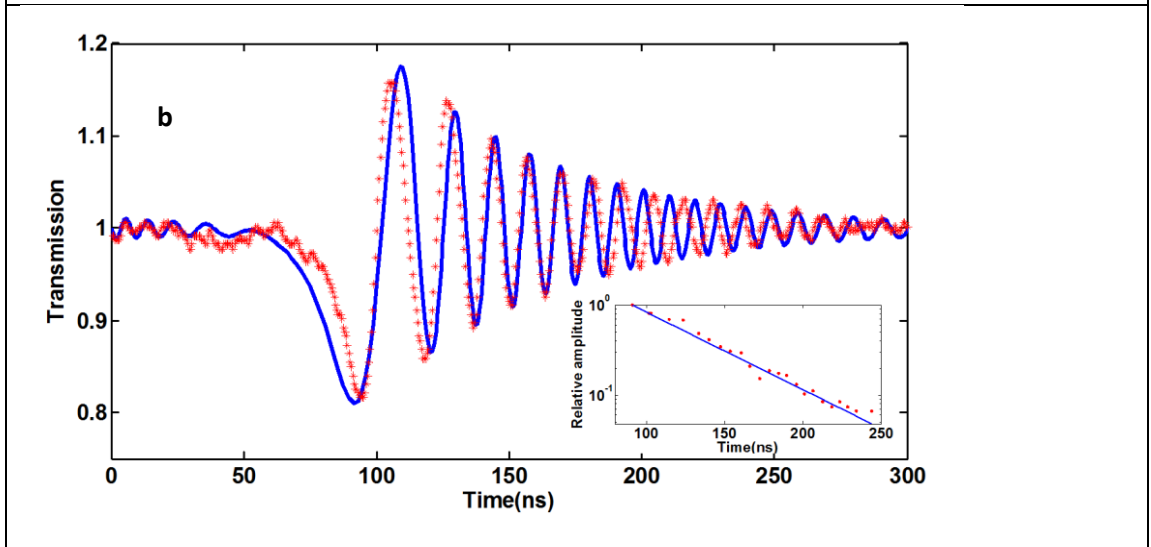

Figure 2.(**a**) Experimental set-up: A tunable laser is evanescently coupled, via a tapered fiber, to an optical whispering-gallery mode (red ring) of a drop resonator. The blue cylinder represents a silica stem that holds the transparent oil.

(b) Experimental results: Ring down signal from a 60 μm oil droplet (red points) together with the best fit to what is predicted by the rate equation (blue line). Inset we plot the peaks of the signal against time on semilogarithmic scale, and used linear regression for extracting the life time of the photon, the calculated quality factor is $Q = (8.5 \pm 0.5) \cdot 10^7$.

When the laser power was above 100 µW and while passing through the droplet optical resonances, strong oscillations in transmission were observed indicating mechanical vibrations. The appearance of these vibrations was so abundant that it was almost impossible to pass through an optical resonance without seeing oscillations as shown in Figure 3.

We experimentally test several droplets (Figure 3a). The droplet diameter ranges between 30 and 100 µm. The droplets micrographs are shown in Figure 3a right next to their Comsol simulated acoustical mode that was 3-5% near the experimentally measured ones. The measured oscillatory signal is shown in Figure 3b. Excitation of high order mechanical modes, as seen here, is typical in optomechanical excitation[12] and relates to the fact that the acoustical cycle time should be of the same scale of the photon lifetime that excites it. When looking at the Fourier spectrum of the oscillations, higher harmonics were seen (Figure 3c) as typical for such oscillations and relates to the fact that optical transmission is nonlinear with radius.

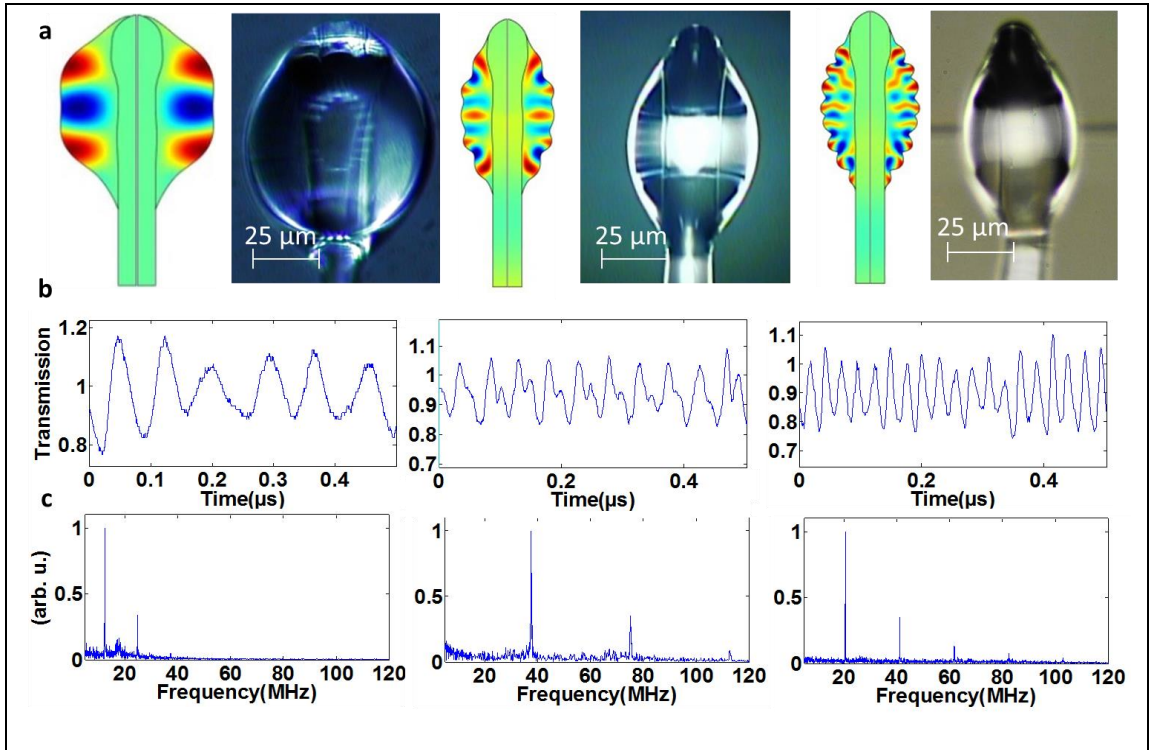

Figure 3 Experimental results: acoustically resonating droplets **a**) Droplet's micrograph near one of its calculated acoustical modes. The droplets diameter is 75 μm, 50 μm and 40 μm. **b**) Oscillations signal for each drop in time domain. **c**) Oscillations signal for each drop in frequency domain where up to 5 harmonics are seen.

The mechanical quality factor of the oil droplets is obtained from the width of the Lorentzian fitted power spectrum of the resonance near oscillation threshold (Figure 4a). Mechanical quality factor for this droplet (Qm), was measured to be $81 \pm 5$. As expected[29], power increment was accompanied by narrowing of the oscillation linewidth from $253\pm15$ KHz at 28 micro watt to $72\pm5$ KHz at 34 microwatt. The power input vs vibration amplitude is given in Figure 4b, and possesses a knee behavior. The location of the knee represents the point where the mechanical gain turns larger than loss and is generally referred to as oscillation threshold. The region below threshold represents modulation of incoming light by thermal Brownian fluctuation while the region above it represents vibration that is optically excited. Threshold was measured this way to be 68 µW.

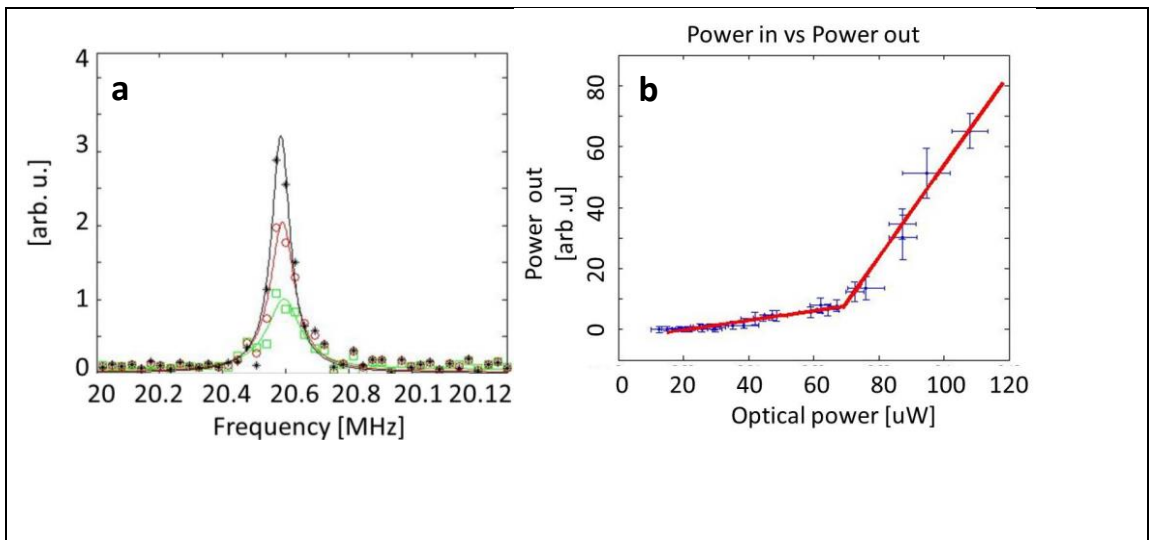

Figure 4 Experimentally **measured mechanical quality factor and threshold a**) Lorentzian fits to the mechanical power spectrum at 3 input powers, green red and black corresponds to $28\pm0.01$µW, $31\pm0.01$µW and $34\pm0.01$µW with linewidth of $253\pm15$KHz, $126\pm9$ KHz and $72\pm5$ KHz respectively. Droplet diameter was 60 µm. **b**) Power in vs power out graph and linear fits for under and over threshold regimes, indicating 68 µW threshold. Droplet diameter was 40µm

In conclusion, we show acoustical resonances in droplets for the first time. Furthermore, our work extends optomechanics to one of the most basic structures in Nature - droplets. Our system is simple as it operates at room condition while fed by a continuous in time laser with no external modulation. Also, no controller or feedback loop is needed in our experiment. These droplets have smooth walls while being simple to fabricate and sustain.

References:

[1] L. Rayleigh, in *Proc. R. Soc. London*1879), pp. 71.
[2] H. Lamb, Proceedings of the London Mathematical Society **1**, 189 (1881).
[3] T. Carmon, H. Rokhsari, L. Yang, T. J. Kippenberg, and K. J. Vahala, Physical Review Letters **94**, 223902 (2005).
[4] O. Arcizet, P.-F. Cohadon, T. Briant, M. Pinard, and A. Heidmann, Nature **444**, 71 (2006).
[5] A. Naik, O. Buu, M. LaHaye, A. Armour, A. Clerk, M. Blencowe, and K. Schwab, Nature **443**, 193 (2006).
[6] D. Kleckner and D. Bouwmeester, Nature **444**, 75 (2006).
[7] S. Gigan *et al.*, Nature **444**, 67 (2006).
[8] A. Jayich, J. Sankey, B. Zwickl, C. Yang, J. Thompson, S. Girvin, A. Clerk, F. Marquardt, and J. Harris, New Journal of Physics **10**, 095008 (2008).
[9] M. Tomes, K. J. Vahala, and T. Carmon, Optics express **17**, 19160 (2009).
[10] M. Eichenfield, J. Chan, R. M. Camacho, K. J. Vahala, and O. Painter, Nature **462**, 78 (2009).
[11] G. Bahl, J. Zehnpfennig, M. Tomes, and T. Carmon, Nature communications **2**, 403 (2011).
[12] T. Carmon and K. J. Vahala, Physical review letters **98**, 123901 (2007).

[13]     M. Tomes and T. Carmon, Physical review letters **102**, 113601 (2009).
[14]     F. Vollmer and S. Arnold, Nature methods **5**, 591 (2008).
[15]     X. Fan and I. M. White, Nature photonics **5**, 591 (2011).
[16]     F. Vollmer and L. Yang, Nanophotonics **1**, 267 (2012).
[17]     Y. Fainman, L. P. Lee, D. Psaltis, and C. Yang, *Optofluidics: fundamentals, devices, and applications* (McGraw-Hill New York:, 2010).
[18]     G. Bahl, K. H. Kim, W. Lee, J. Liu, X. Fan, and T. Carmon, Nature communications **4** (2013).
[19]     K. H. Kim, G. Bahl, W. Lee, J. Liu, M. Tomes, X. Fan, and T. Carmon, Light: Science & Applications **2**, e110 (2013).
[20]     A. Ashkin and J. Dziedzic, Physical Review Letters **38**, 1351 (1977).
[21]     H.-M. Tzeng, K. F. Wall, M. Long, and R. Chang, Optics letters **9**, 499 (1984).
[22]     M. Hossein-Zadeh and K. J. Vahala, Optics express **14**, 10800 (2006).
[23]     A. Jonáš, Y. Karadag, M. Mestre, and A. Kiraz, JOSA B **29**, 3240 (2012).
[24]     P.-G. De Gennes, F. Brochard-Wyart, and D. Quéré, *Capillarity and wetting phenomena: drops, bubbles, pearls, waves* (Springer Science & Business Media, 2004).
[25]     J. Knight, G. Cheung, F. Jacques, and T. Birks, Optics letters **22**, 1129 (1997).
[26]     A. A. Savchenkov, A. B. Matsko, V. S. Ilchenko, and L. Maleki, Optics Express **15**, 6768 (2007).
[27]     T. Carmon, L. Yang, and K. Vahala, Optics Express **12**, 4742 (2004).
[28]     M. L. Gorodetsky and V. S. Ilchenko, JOSA B **16**, 147 (1999).
[29]     H. Rokhsari, T. Kippenberg, T. Carmon, and K. Vahala, Selected Topics in Quantum Electronics, IEEE Journal of **12**, 96 (2006).